\documentclass{INTERSPEECH2023}
\usepackage[ruled,linesnumbered]{algorithm2e}
\SetKwInput{KwInput}{Input} 
\SetKwInput{KwOutput}{Output}
\usepackage{xcolor}
\usepackage{color}
\usepackage{graphicx}
\usepackage{subcaption}
\usepackage{multirow}
\usepackage{cite}

\SetCommentSty{mycommfont}

\interspeechcameraready

\title{CoMFLP: Correlation Measure based Fast Search on ASR Layer Pruning}
\name{Wei Liu, Zhiyuan Peng, Tan Lee}
\address{
  Department of Electronic Engineering, The Chinese University of Hong Kong 
  }
\email{louislau\_1129@link.cuhk.edu.hk, jerrypeng1937@gmail.com, tanlee@cuhk.edu.hk}

\begin{document}
\maketitle
\begin{abstract}
Transformer-based speech recognition (ASR) model with deep layers exhibited significant performance improvement. However, the model is inefficient for deployment on resource-constrained devices.
Layer pruning (LP) is a commonly used compression method to remove redundant layers. Previous studies on LP usually identify the redundant layers according to a task-specific evaluation metric. They are time-consuming for models with a large number of layers, even in a greedy search manner. To address this problem, we propose CoMFLP, a fast search LP algorithm  based on correlation measure. 
The correlation between layers is computed to generate a correlation matrix, which identifies the  redundancy among layers. The search process is carried out in two steps: (1) coarse search:  to determine top $K$ candidates by pruning the most redundant layers based on the correlation matrix; (2) fine search: to select the best pruning proposal among $K$ candidates using a task-specific evaluation metric. Experiments on an ASR task show that the pruning proposal determined by CoMFLP outperforms existing LP methods while only requiring constant time complexity. The code is publicly available at 
\url{https://github.com/louislau1129/CoMFLP}.
\end{abstract}

\section{Introduction}



Neural network based ASR models with deep layers have shown superior performance \cite{li2022recent, zhang2022bigssl, wang2022improving}. This is mainly due to the recent advances in the transformer architecture \cite{DBLP:conf/interspeech/GulatiQCPZYHWZW20, chen2020non, vaswani2017attention} and the effectiveness of scaling laws \cite{kaplan2020scaling}. 
However, large and deep models are difficult, if not impossible, to be deployed on resource-constrained devices at the edge \cite{vipperla2020learning,shangguan2019optimizing, lin2018edgespeechnets}.

Model compression \cite{choudhary2020comprehensive, guo2018survey, gou2021knowledge, hinton2015distilling, augasta2013pruning, wang2019cop, blalock2020state} was extensively studied for realizing large models on small-footprint devices without significant performance degradation. One of the predominant approaches is pruning \cite{augasta2013pruning, wang2019cop, blalock2020state}. 
Depending on whether the pruned parameters are grouped according to inherent structures (e.g., channel, head, and layer), there are unstructured \cite{chen2020lottery} and structured pruning \cite{michel2019sixteen, mccarley2019structured, sajjad2020poor, DBLP:conf/iclr/FanGJ20} approaches.  Unstructured pruning treats each individual weight as the unit of pruning, often resulting in a sparse model that is not suitable for speedup on current hardware\cite{sanh2020movement}. Structured pruning can achieve great acceleration, especially when dropping many layers \cite{DBLP:conf/iclr/FanGJ20, DBLP:conf/acl/XiaZC22}.


It is noted that the transformer stacked with a large number of layers, e.g., DeepNet with 1,000 layers \cite{wang2022deepnet}, demonstrates significant performance. However, it comes at the cost of higher computational resources and memory consumption. 
It is hypothesized that some of the many layers might be redundant and have little contribution to the overall system performance \cite{DBLP:conf/interspeech/LeeK021}. This motivates the present study to inspect the redundancy among layers and perform layer-level structured pruning, i.e., layer pruning (LP), for simplifying deep models. 



Let $L$ be the total number of layers. The core of LP is a pruning proposal that stipulates $N$ of the $L$ layers to be removed.
The optimal pruning proposal can be found by enumerating all possible combinations of  $N$ layers and selecting the best one. For ASR systems, performance metrics like word error rate (WER) are used for determining the best proposal. Apparently, the high time complexity $\mathcal{O}(C_{L}^{N})$ makes exhaustive search practically infeasible.  In \cite{chen2018shallowing},  the method of linear probing was applied to estimate the importance of different layers. The layer output was extracted as an input feature to a linear classifier for task-specific performance evaluation. If two consecutive layers' performances are similar, one of them would be removed. In this case, a total of $L$ classifiers need to be trained.  In \cite{peer2022greedy},  greedy layer pruning (GLP) was proposed to reduce the complexity of the pruning proposal search. The key idea is that
the best solution for pruning $(n+1)$ layers is built on top of the already known best solution for pruning $n$ layers, such that only $(L-n)$ possible combinations need to be evaluated.  With this simplification, the complexity of the pruning proposal search is reduced from $\mathcal{O}(C_{L}^{N})$ to $\mathcal{O}(L*N)$. Several other pruning strategies, like top-layer, middle-layer, and bottom-layer dropping, etc., were empirically studied in \cite{sajjad2023effect}.

\begin{figure*}[ht!]
    \vspace{-3mm}
  \centering
  \includegraphics[width=0.75\linewidth]{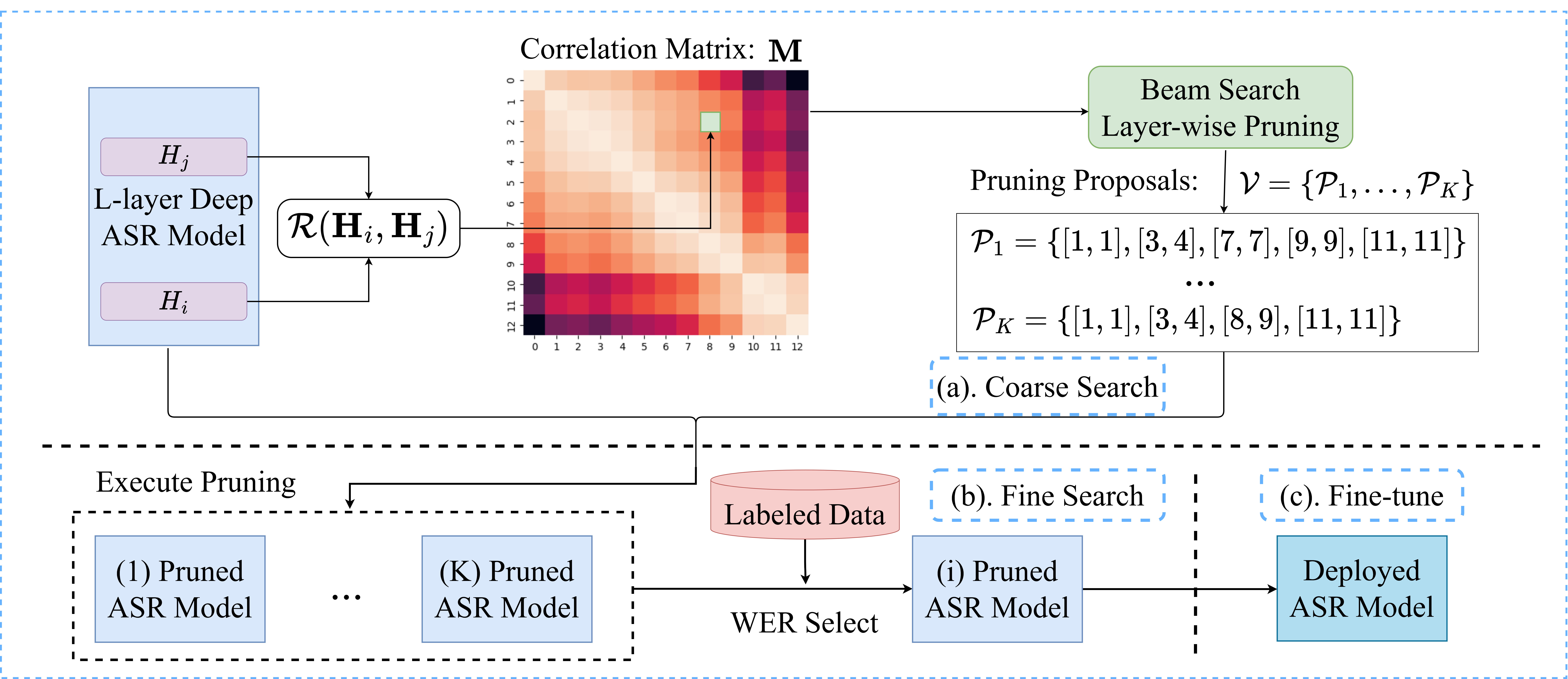}
  \caption{The overall diagram of the proposed CoMFLP method on ASR, where $L$=12 is illustrated as an example.}
  \label{fig:lp_diagram}
\end{figure*}



Inspired by \cite{DBLP:conf/interspeech/LeeK021}, if two layers' outputs are similar, the layers between them are assumed to be redundant and can be discarded. In the present study, we propose the \textbf{Co}rrelation \textbf{M}easure based \textbf{F}ast Search on \textbf{L}ayer \textbf{P}runing (CoMFLP). Two correlation methods, singular vector canonical correlation analysis (SVCCA) and distance correlation (DC), are computed for any pair of layers, leading to a correlation matrix that represents the pair-wise layer similarity.  Specifically, CoMFLP comprises two steps; (1) coarse search: top-$K$ pruning proposals are determined using a layer-wise beam search on the correlation matrix; (2) fine search: WER is used to select the best one among the $K$ candidates. The time complexity of the proposed method is approximately constant $\mathcal{O}(K)$ since step (1) has negligible computation cost. It is much faster than WER-based GLP ($\mathcal{O}(L*N)$) and has a better performance. Experiments show that the pruning proposal determined by CoMFLP can provide a very promising starting point for later fine-tuning in terms of convergence and initialization.

\section{CoMFLP}

We inspect a well-trained neural network for ASR and attempt to apply layer pruning  to achieve good compromise between efficiency and performance. 
A two-stage fast search algorithm is proposed. As depicted in Fig. \ref{fig:lp_diagram}, it includes a coarse search and a fine search. Given an $L$-layer network, an $(L+1)$-by-$(L+1)$ correlation matrix is generated in the coarse search. 
With the correlation matrix, pruning proposals are evaluated, compared, and ranked. The top $K$ proposals $\mathcal{V} = \{\mathcal{P}_1, ..., \mathcal{P}_K\}$ are retained for the subsequent fine search. In the fine search, task-specific performance metrics on labeled data are used to evaluate the $K$ candidate proposals and select the best one. Only constant time complexity $\mathcal{O}(K)$ is required for the entire search.

\subsection{Correlation matrix}
\label{subsec:corr_matrix}
The correlation matrix measures the similarity among layers. If the input and output of a block of layers are very similar, the layers within that black are considered to be redundant. Layer redundancy thus can be reflected by the correlation matrix.
For speech recognition, consider a batch of $B$ speech samples $\mathbf{S} = \{\mathbf{s}_i\}_{i=1}^{B}$. In a  CTC-based ASR system, the samples are zero-padded and encoded into frame-level hidden representations $\mathbf{H}_{0} \in \mathbb{R}^{B \times T_{0} \times D_{0}}$, where $T_{0}$ denotes the number of frames and $D_0$ represents the hidden dimension. The hidden representation $\mathbf{H}_0$ is then forwarded into an $L$-layer Transformer $\mathbf{Enc}$. The layer-wise forward process can be formulated as follows:

\begin{equation}
    \mathbf{H}_i = \mathbf{Enc}^{i} (\mathbf{H}_{i-1}); ~~ 1 \leq i \leq L,
\end{equation}
where $\mathbf{H}_{i} \in \mathbb{R}^{B \times T_{i} \times D_i}$ denotes the hidden representation produced by the $i$-th layer of the transformer $\mathbf{Enc}$. The hidden representations are averaged along the time axis, i.e.,
$\mathbf{\Bar{H}}_i \longleftarrow \text{TempoAverage}(\mathbf{H}_{i})$, where $\Bar{\mathbf{H}}_i \in \mathbb{R}^{B \times D_i}$.
Finally, singular vector canonical correlation analysis (SVCCA) or distance correlation (DC) is applied on $\{\mathbf{\Bar{H}}_i\}_{i=0}^L$ to derive the correlation matrix $\mathbf{M} \in \mathbb{R}^{(L+1)\times(L+1)}$ .

\subsubsection{SVCCA}
\label{subsec:svcca}
SVCCA is developed to measure the similarity between two hidden layer representations \cite{raghu2017svcca}. It involves two steps, namely singular value decomposition (SVD) and canonical correlation analysis (CCA)\cite{bach2005probabilistic}. SVD is applied to $\mathbf{\Bar{H}}_i$ to obtain the subspace $\mathbf{\Bar{H}}_{i}^{\prime} \in \mathbb{R}^{B \times D_{i}^{\prime}}$ that corresponds to the $D_{i}^{\prime}$ largest singular values (up to a pre-defined ratio of the total variance, e.g., 99\%). Similarly, $\mathbf{\Bar{H}}_{j}^{\prime} \in \mathbb{R}^{B \times D_{j}^{\prime}}$ can be obtained by applying SVD to $\mathbf{\Bar{H}}_{j}$ .
Second, CCA is applied to project $\mathbf{\Bar{H}}_{i}^{\prime}$ and $\mathbf{\Bar{H}}_{j}^{\prime}$ into a shared subspace, where the correlation between the projected matrices is maximized. The correlation coefficients are averaged into a scalar in the range of 0 to 1, which is defined as the SVCCA similarity score between $\Bar{\mathbf{H}}_i$ and $\Bar{\mathbf{H}}_j$. 
A detailed description of SVCCA can be found in \cite{raghu2017svcca}. Applying SVCCA on all possible $i,j$ gives the correlation matrix $\mathbf{M}$.

\subsubsection{DC}
DC \cite{zhen2022versatile, szekely2007measuring} provides a statistical measure of dependence between random vectors.
Consider the hidden representation $\mathbf{\Bar{H}}_i \in \mathbb{R}^{B \times D_i}$, where the row vector $\mathbf{h}_k^{i} \in \mathbb{R}^{D_i}$ corresponds to the $i$-th layer's response of the $k$-th sample. The distance matrix $\mathbf{A}^i \in \mathbb{R}^{B\times B}$ measures the normalized distances between row vectors in $\mathbf{\Bar{H}}_i$. It can be computed as follows,

\begin{align}
\label{eq:distance_matrix}
    \mathbf{a}_{k,l}^{i} &= ||\mathbf{h}_k^{i} - \mathbf{h}_l^{i}||_{2}, \\
\label{eq:normalize_dist_mat}
    \mathbf{A}_{k,l}^i &= \mathbf{a}_{k,l}^{i} - \frac{1}{M}\sum_{l=1}^{M} \mathbf{a}_{k,l}^{i} - \frac{1}{M}\sum_{k=1}^{M} \mathbf{a}_{k,l}^{i} + \frac{1}{M^2}\sum_{k,l=1}^{M} \mathbf{a}_{k,l}^{i},
\end{align}
where $\mathbf{A}_{k,l}^i$ refers to the element of $\mathbf{A}^i$ at $k$-th row and $l$-th column. To this end, the similarity score $\mathbf{M}_{i,j}$ between $\Bar{\mathbf{H}}_i$ and $\Bar{\mathbf{H}}_j$ is given by,

\begin{equation}
    \label{eq:dc}
    \mathbf{M}_{i,j} = (\frac{V_{i,j}}{\sqrt{V_{i,i}V_{j,j}}} )^{\frac{1}{2}}, 
\end{equation}
where the distance covariance $V_{i,j} = \frac{1}{B^{2}}\sum_{k,l=1}^{B}\mathbf{A}^i_{k,l}\mathbf{A}^j_{k,l}$ and $ 0 \leq \mathbf{M}_{i,j} \leq 1$. 

\subsection{Coarse search}


The coarse search is to efficiently find out (maybe with loss of precision) the top $K$  pruning proposals.
To perform a coarse search, a method of  measuring the quality of a pruning proposal is needed. In the correlation matrix $\mathbf{M}$, the element $\mathbf{M}_{i,j}$ represents the similarity between the outputs from the $i$-th layer and the $j$-th layer. It reflects the redundancy of layers $i+1$ to $j$, and therefore can be leveraged to measure the quality of pruning proposals. Consider a pruning proposal $\mathcal{P} = \{[s_1, e_1], ..., [s_p, e_p]\}$ that discards the layers from $s_1$ to $e_1$,$\cdots$, $s_p$ to $e_p$, where $1 \leq s_1 \leq e_1 < \cdots < s_p \leq e_p \leq L$. The quality of $\mathcal{P}$ is defined as:
\begin{equation}
\label{eq:val}
    v(\mathcal{P}) = \frac{1}{p}\sum_{t=1}^{p} \mathbf{M}_{s_t -1, e_t}.
\end{equation}
$v(\mathcal{P})$ is expected to positively correlate with the performance of the pruned model. Higher quality would likely (but not necessarily \footnote{
E.g., SVCCA is invariant to affine transformation, which cannot be detected by correlation but may have a side effect on the performance.
}) lead to a better pruned model. 

Consider the goal of pruning $N$ layers from an $L$-layer neural network. There could be numerous pruning proposals, making the exact search for the best one, if not intractable, time-consuming. On the other hand, the coarse search is able to efficiently find out the top $K$ pruning proposals that can be regarded as the most potential candidates. The intuition is that a good $N$-layer pruning proposal is most likely to be derived from that of pruning $(N-1)$ layers. This leads to a recursive beam search algorithm, as illustrated in Algorithm \ref{alg:beam_search}. Specifically, suppose that a set of $K$ proposals of pruning $(N-1)$ layers are found. For each of the proposals, we enumerate all possible choices of selecting an additional layer for pruning, leading to new proposals of pruning $N$ layers. The new proposals are ranked according to the metric defined in Eq. (\ref{eq:val}). Finally, only the top $K$ good proposals are selected. Algorithm \ref{alg:beam_search} provides the implementation details, where a \textit{min-queue} is used for fast ranking of the top $K$ good proposals.

\vspace{-2mm}
\begin{algorithm}
\caption{function \textbf{coarse\_search}($L, N, K$)}
\label{alg:beam_search}
\textbf{Notation} {$L$: \# layers; $N$: \# layers to be pruned; $K$: beam size;} \\
\KwOutput{$K$ pruning proposals: $\mathcal{V} = \{ \mathcal{P}_1, \cdots, \mathcal{P}_K\}$}
\textbf{Init} $\mathcal{V} = \varnothing$; $\mathcal{U} = \varnothing$\\
\tcc{A min priority queue that keeps the K largest elements}
\textbf{Init} $\mathcal{L}$ as an \textit{\textbf{min-queue}} filled with $K$ tuples $(0, \varnothing)$ 
\If{ $N$ is 0}{ \textbf{Return} $\{\varnothing \}$}
\For{$\mathcal{P} \in$ \textnormal{\textbf{coarse\_search}}($L, N-1, K$)}{
    \For{ layer $l \in [1, L]\setminus\mathcal{P}$}{
        $\mathcal{P}^* = \mathcal{P}\cup [l, l]$ \tcp*[r]{add a new layer to prune} 
        $v^* = v(\mathcal{P}^*)$  \tcp*[r]{the quality metric, refer to Eq. (\ref{eq:val})}
        \If{$\mathcal{P}^* \notin \mathcal{U}$}
        {
        $\mathcal{U} = \mathcal{U} \cup \{\mathcal{P}^*\}$  \tcp*[r]{deduplicate the set of $\mathcal{P}^*$ }
        \If{$v^* > \min(\mathcal{L})$}{
            pop($\mathcal{L}$) \tcp*[r]{remove the tuple with smallest $v^*$ } 
            push($\mathcal{L}, (v^*, \mathcal{P}^*$)) \tcp*[r]{add the new tuple}
        } 
        }       
    }
}
\For{$(v, \mathcal{P}) \in \mathcal{L}$}{
    $\mathcal{V} = \mathcal{V} \cup \{\mathcal{P}\}$
}
\textbf{Return} $\mathcal{V}$
\end{algorithm}

\vspace{-2mm}
\subsection{Fine search}
\label{subsec:fine search}
The fine search aims to select the best candidate from the $K$ pruning proposals generated by the coarse search. To achieve this goal, we execute the pruning proposals to generate $K$ pruned models and evaluate them on labeled data. Word error rate (WER) is used as the evaluation metric for ranking. To this end, the best pruned model is selected and fine-tuned. The underlying hypothesis is that a pruned model with a lower WER could generally speed up the fine-tuning process and is very likely to achieve better performance.




\section{Experimental Setup}
\subsection{Model, data, and metric}

Two Wav2vec2-CTC models \cite{DBLP:conf/slt/LuC22} from Huggingface are leveraged for layer pruning in our experiments. The first one, denoted as W2V2CTC-12\footnote{kehanlu/mandarin-wav2vec2-aishell1}, is a 12-layer transformer and pre-trained on AISHELL-2\cite{du2018aishell}. The second one, named as W2V2CTC-24\footnote{qinyue/wav2vec2-large-xlsr-53-chinese-zn-cn-aishell1}, is a 24-layer transformer and based on \textit{wav2vec2-large-xlsr-53} \cite{DBLP:conf/interspeech/ConneauBCMA21}. Both two models are fine-tuned on AISHELL-1 \cite{bu2017aishell}. 
The \textit{dev} and \textit{test} sets of AISHELL-1 are utilized for evaluation. Character error rate (CER) is used as the evaluation metric since our ASR outputs are Chinese characters.

\subsection{Configuration for search}
\begin{itemize}[leftmargin=*]
    \item \textbf{Correlation matrix: } 
    In SVCCA, the batch size $B$ should be at least 5-10 times larger than the layer dimension $D_i$ for accurate estimation of the correlation matrix \cite{raghu2017svcca}. In DC, the batch size $B$ can be small. However, it requires multiple batches (\textit{num\_batch}) of samples to estimate. The estimated correlation matrices are averaged as the final result. 
    Our preliminary experiments suggest the hyper-parameters as follows: (1) SVCCA: $B=9600$, 99\% variance kept in SVD; (2) DC: $B=4$, \textit{num\_batch}$=10$. Note that the speech samples are drawn from the training set of AISHELL-1.
    
    
    \item \textbf{Coarse search: } the beam size $K$ in Algorithm \ref{alg:beam_search} is set to $10$. 

    \item \textbf{Fine search: } 320 utterances are randomly selected from the \textit{test} set and denoted as the \textit{valid} set. They are used for decoding and computing CER in fine search.
\end{itemize}

\subsection{Configuration for fine-tuning}
The best pruned model is fine-tuned on the training set of AISHELL-1 for 50 epochs. AdamW \cite{DBLP:conf/iclr/LoshchilovH19} is adopted as the optimizer with an initial learning rate of 2e-5. The scheduler \textit{ReduceLROnPlateau} is applied with \textit{patience} = 10 and \textit{factor} = 0.5. Other hyperparameters are set as follows: \textit{batch\_size = 32}, \textit{freeze\_feature\_extractor = False}, \textit{clip\_grad\_norm = None}, and \textit{layerdrop = 0.1}. No data augmentation is involved.


\section{Results and Analysis}


\begin{figure}[ht]
  \centering
  \subfloat[W2V2CTC-12 clip12to6]
   {\includegraphics[width=0.25\textwidth]{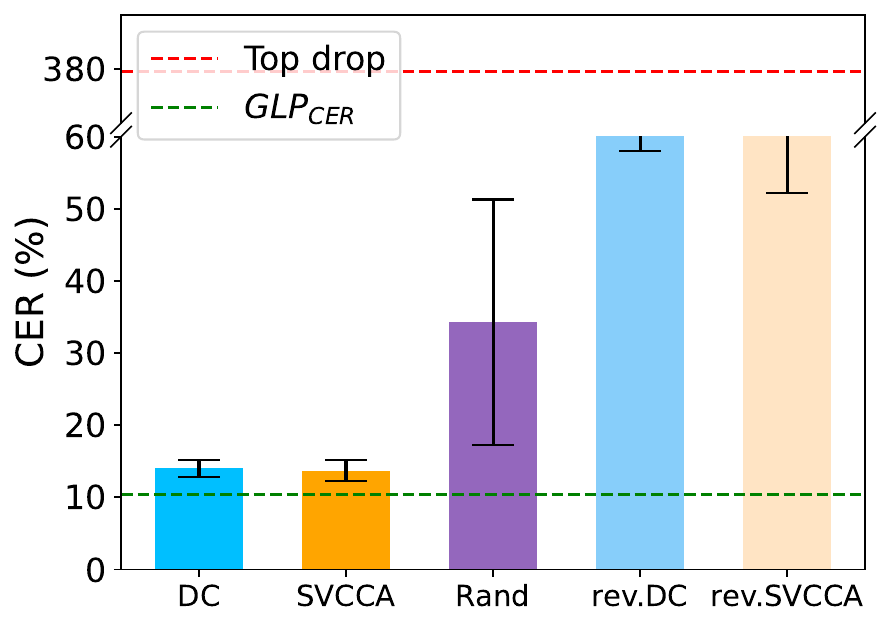}}
  \subfloat
    [W2V2CTC-12 pruning]
   {\includegraphics[width=0.25\textwidth]{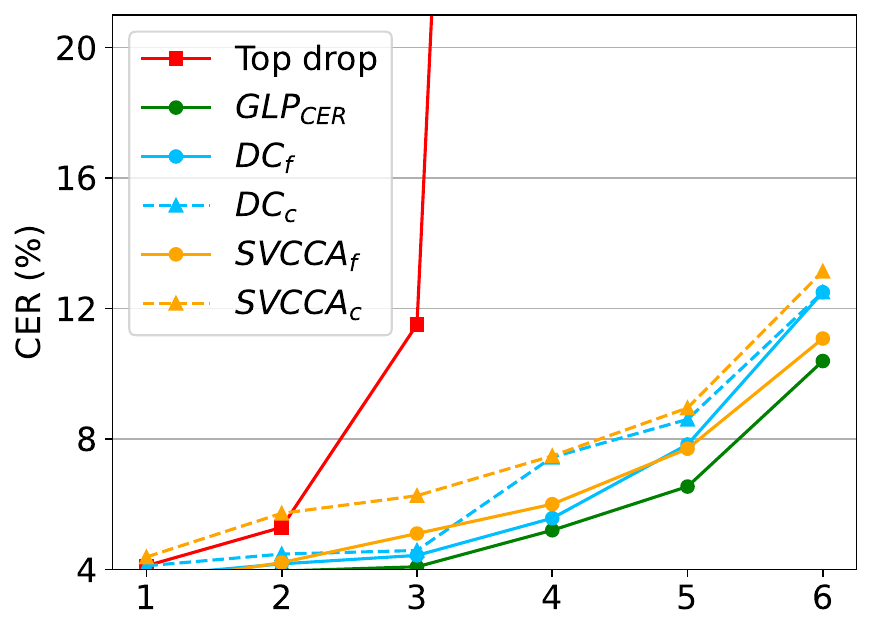}}
   
  \subfloat
    [W2V2CTC-24 clip24to12]
   {\includegraphics[width=0.25\textwidth]{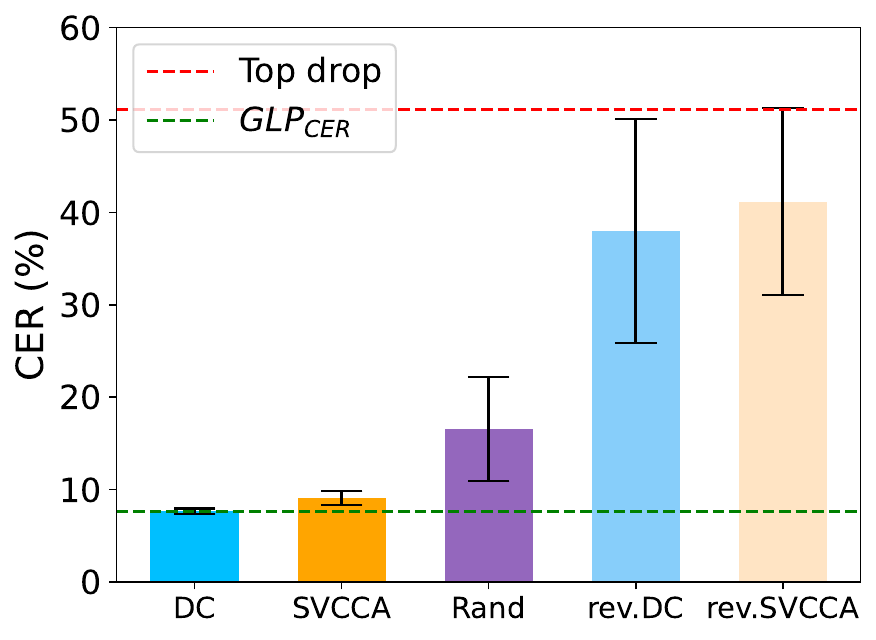}}
  \subfloat
    [W2V2CTC-24 pruning]
   {\includegraphics[width=0.25\textwidth]{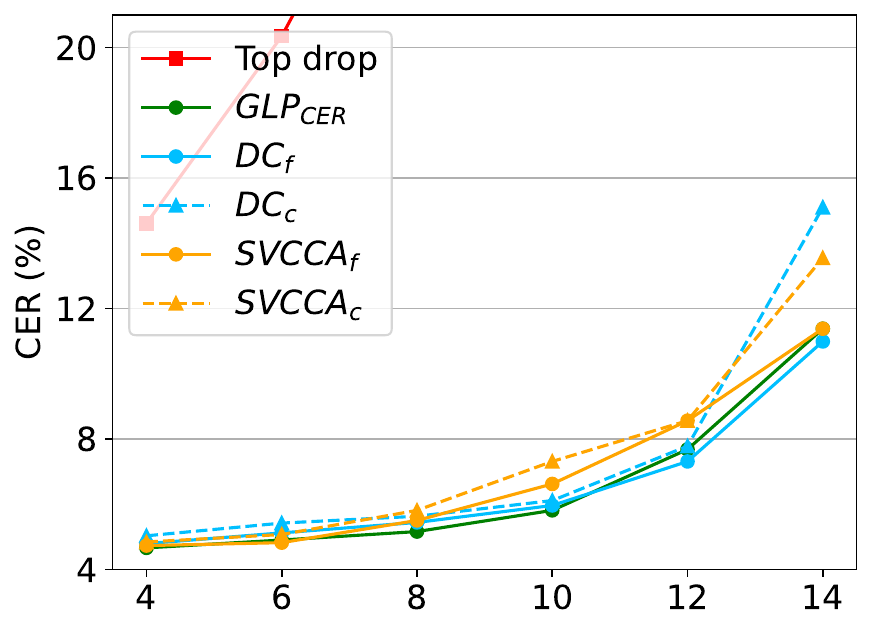}}

  \caption{The CER results on \textit{valid} set of different layer pruning methods. (a) and (b) are experimented on the W2V2CTC-12 model, while (c) and (d) are for the W2V2CTC-24 model.}
\label{fig:effect_of_CoMFLP}
\vspace{-3mm}
\end{figure}

\subsection{Performance evaluation without fine-tuning}
The CER results on the $valid$ set are used to assess the effect of layer pruning. Fig. \ref{fig:effect_of_CoMFLP} compares the performance of various pruning methods. 
In Fig. \ref{fig:effect_of_CoMFLP} (a) and (c), \textit{DC/SVCCA} give the coarse search results. The bar height represents the CER mean of the searched 10 pruning proposals, and the error bar denotes their standard deviation (\textit{std.}). 
The prefix ``rev.'' before \textit{DC/SVCCA} means that the correlation-based search is reversely performed, where the worst pruning proposals are given. \textit{Rand} is to randomly select 10 pruning proposals. Comparing \textit{DC/SVCCA} bars with other bars, the correlation-based search CoMFLP is found to be able to distinguish the good candidates from those terrible ones. 
In addition, two horizontal lines are given as references. \textit{GLP\textsubscript{CER}} is a powerful  approach that directly searches with CER in a greedy manner, achieving a relatively good balance between speed and performance. The Top drop is performed by simply dropping the top several layers.  It can be observed that the \textit{DC/SVCCA} bars exhibit comparable performance with \textit{GLP\textsubscript{CER}} (green line) and are significantly superior to the \textit{Top drop} (red line).

Fig. \ref{fig:effect_of_CoMFLP} (b) and (d) give the performance trend with the varied numbers of pruned layers. When dropping more layers, all methods show monotonically increasing CERs. The subscript ``f" denotes the best pruning result after the fine search, while the ``c" represents only the coarse search 
 performed, and the pruning proposal indicated by the highest quality metric is output. The \textit{DC\textsubscript{c}} and \textit{SVCCA\textsubscript{c}} are shown to perform worse than their counterparts with a fine search. 

In conclusion, the proposed CoMFLP exhibits great power to propose a good pruning proposal that is comparable to \textit{GLP\textsubscript{CER}} while requiring only a constant time complexity. 

\subsection{Performance evaluation after fine-tuning}
We then analyze the performance of the pruned model after fine-tuning (FT). Experiments are performed on the pruned W2V2CTC-12 (\textit{clip12to6}). The results of different pruning strategies are given in Table \ref{tab:ft_result}.
The CoMFLP series achieve the lowest CER results, significantly better than \textit{Scratch} and \textit{Top drop}. 
The \textit{GLP\textsubscript{CER}} achieves only 6.62\% CER on \textit{test} set despite the best performance before FT (cf. Fig. \ref{fig:effect_of_CoMFLP} (a)). Its low quality in contrast to CoMFLP (0.93 vs. 0.96) could be the reason.

In the CoMFLP block, \textit{DC\textsubscript{f}} and \textit{SVCCA\textsubscript{f}} do not guarantee a better FT result. This may be due to the rather small CER \textit{std.} shown in the current coarse search results (cf. Fig. \ref{fig:effect_of_CoMFLP} (a)). The fine search serves to filter out potentially bad proposals that are of high correlation but terrible CER. It is worth noting that \textit{SVCCA\textsubscript{c}} is exactly the proposal to drop every other layer, which is quite reasonable and intuitive for the \textit{clip12to6} case. 

Note that LP can be complementary to other compression methods. Better FT results can be achieved by combining the knowledge distillation from the original deep model \cite{chen2021knowledge}.


\begin{table}[t]
\centering
\caption{The fine-tuning CER (\%) results of different models on \textit{dev} and \textit{test} sets of AISHELL-1. The last column  represents the pruning proposal's quality metric measured by DC.}
\label{tab:ft_result}
\begin{tabular}{ccccc}
\toprule
\multicolumn{2}{c|}{Model}                                                       & dev  & \multicolumn{1}{c|}{test}  & quality \\ \midrule
\multicolumn{2}{c|}{W2V2CTC-12}                                                  & 5.10 & \multicolumn{1}{c|}{5.44}  & -      \\ \hline
\multicolumn{5}{c}{Layer pruning to 6 layers}                                                                                  \\ \midrule
\multicolumn{2}{c|}{\textit{Scratch}}                                                     & 9.04 & \multicolumn{1}{c|}{10.11} & -      \\
\multicolumn{2}{c|}{\textit{Top drop}}                                                    & 7.21 & \multicolumn{1}{c|}{7.80}  & 0.69    \\
\multicolumn{2}{c|}{\textit{GLP\textsubscript{CER}}}                                                & 6.13 & \multicolumn{1}{c|}{6.62}  & 0.93    \\ \midrule
\multicolumn{1}{c|}{\multirow{4}{*}{CoMFLP}} & \multicolumn{1}{c|}{\textit{DC\textsubscript{c}}}    & 5.86 & \multicolumn{1}{c|}{6.28}  & 0.95    \\
\multicolumn{1}{c|}{}                        & \multicolumn{1}{c|}{\textit{SVCCA\textsubscript{c}}} & \textbf{5.76}     & \multicolumn{1}{c|}{\textbf{6.16}}      & \textbf{0.96}         \\ \cline{2-5} 
\multicolumn{1}{c|}{}                        & \multicolumn{1}{c|}{\textit{DC\textsubscript{f}}}    & 5.86 & \multicolumn{1}{c|}{6.28}  & 0.95    \\
\multicolumn{1}{c|}{}                        & \multicolumn{1}{c|}{\textit{SVCCA\textsubscript{f}}} &  5.91    & \multicolumn{1}{c|}{6.41}      &  0.94       \\ \bottomrule
\end{tabular}
\end{table}


\subsection{Training dynamics analysis}
\begin{figure}[h]
  \centering
  \subfloat[CER training curves.]
   {\includegraphics[width=0.245\textwidth]{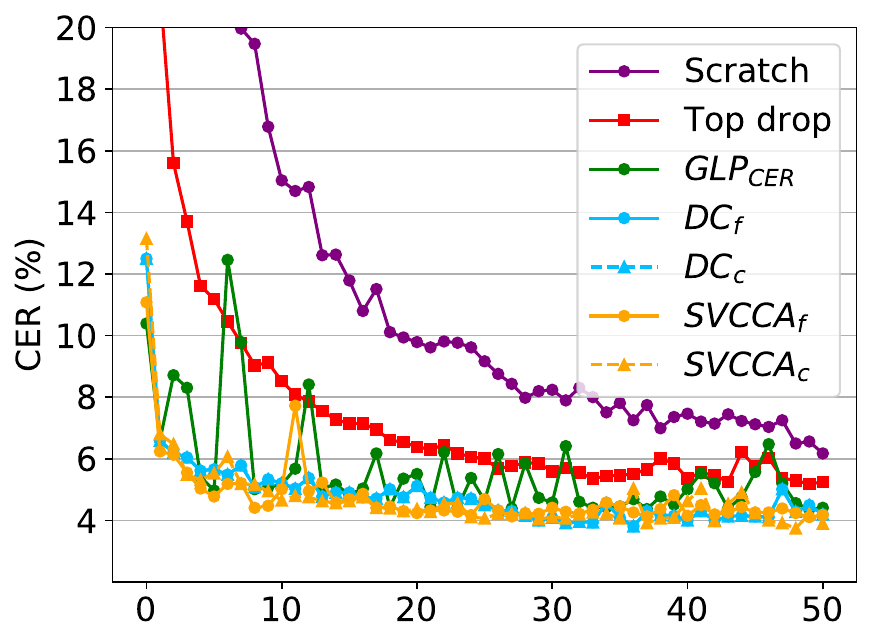}}
  \subfloat[Dynamic correlation matrix.]
   {\includegraphics[width=0.255\textwidth]{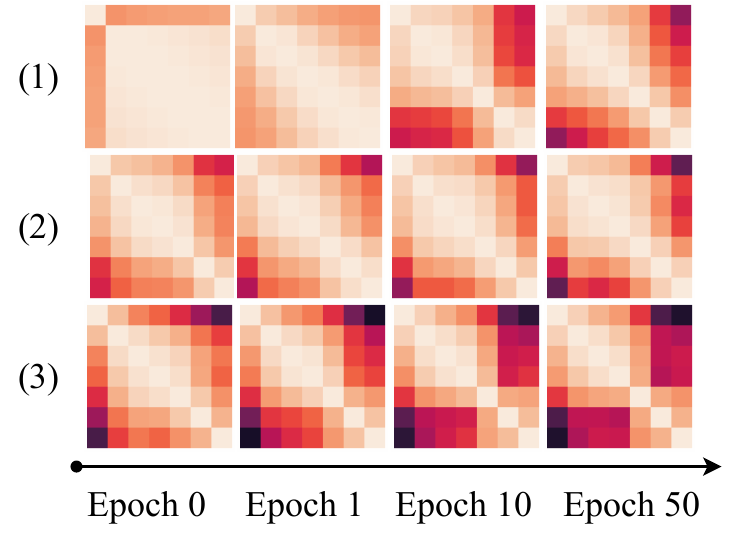}}
  
  \caption{Training Dynamics with (a) CER and (b) Correlation matrix $\mathbf{M}$. Specifically, b.(1)-b.(3) represents three 6-layer models, namely \textit{Scratch}, \textit{Top drop}, and \textit{DC\textsubscript{f}}.}
\label{fig:training_dynamic}
\vspace{-2mm}
\end{figure}

To further illustrate the superiority of CoMFLP, the training dynamics \cite{tirumala2022memorization} are visualized in Fig. \ref{fig:training_dynamic}. It can be clearly seen that in Fig. \ref{fig:training_dynamic} (a), the \textit{DC/SVCCA} series converge faster and are more stable than other curves. The \textit{GLP\textsubscript{CER}} curve is found to have a bit of fluctuation, which may imply its sub-optimal performance. 

In Fig. \ref{fig:training_dynamic} (b), the correlation matrices of three different 6-layer models over the training process are plotted as a 3-by-4 heatmap array. They are (1) \textit{Scratch}, (2) \textit{Top drop}, and (3) \textit{DC\textsubscript{f}}. Each of them gives four epoch snapshots.  For all three groups, the brightness of the heatmaps becomes darker as training progresses, indicating that the layer redundancy of the network tends to be reduced. From the vertical view, we can see that the \textit{DC\textsubscript{f}}'s heatmap is darker than the \textit{Top drop}'s heatmap and further darker than that of the \textit{Scratch}. This could partially explain why the FT performance of \textit{DC\textsubscript{f}} is better than \textit{Top drop}, and the \textit{Scratch} performs the worst. A starting point model at epoch 0 with less redundancy means that different layer representations tend to be more diverse. This initialized diversity could lead to more effective learning during the training process, especially for models with limited capacity. 

\section{Conclusions}
This paper presents a correlation measure based fast search layer pruning method called CoMFLP. Two correlation measures, i.e., DC and SVCCA have been attempted. Only constant time complexity is required for the entire search. Experiments on an ASR task show that CoMFLP can efficiently determine pruning proposals of high quality. When applying the searched pruning proposal, better convergence and initialized diversity are found in the fine-tuning process. 
In future work, the effectiveness and generalization of the method should be examined on deeper layers than the current setup for different tasks. 

\bibliographystyle{IEEEtran}
\bibliography{mybib}

\end{document}